\documentclass[twocolumn,showpacs,preprintnumbers,amsmath,amssymb, floatfix]{revtex4}

\usepackage{epsfig}
\usepackage{graphicx}
\usepackage{bm}
\usepackage{multirow}

\begin{document}


\title{Characterization of Chaotic Motion in a Rotating Drum}

\author{James E.~Davidheiser}%
\email{jamesdavidheiser@gmail.com}
\author{Paul Syers}
\author{P.~N.~Segr\`e}
\author{Eric R.~Weeks}%
\email{weeks@physics.emory.edu}

\affiliation{Physics Department, Emory University, Atlanta, GA 30322}

\date{\today}

\begin{abstract}
Numerous studies have demonstrated the potential
for simple fluid plus particle systems to produce
complicated dynamical behavior.  In this work, we study a horizontal
rotating drum filled with pure glycerol and three large, heavy
spheres.  The rotation of the drum causes the spheres to cascade and
tumble and thus interact with each other.  We find several different
behaviors of the spheres depending on the drum rotation rate.
Simpler states include the spheres remaining well separated, or
states where two or all three of the spheres come together and
cascade together.  We also see two more complex states, where
two or three of the spheres move chaotically.  We characterize
these chaotic states and find that in many respects they are
quite unpredictable.  This experiment serves as a simple model
system to demonstrate chaotic behavior in fluid dynamical systems.
\end{abstract}


\maketitle

\section{ \label{sec:level1}Introduction and Prior Work}

Prior experiments and simulations have shown that particles
interacting with fluids can behave chaotically.  Simple fluid
flows with nonspherical particles \cite{nonspherical,spheroids},
stirred particle-filled fluids \cite{advection}, and various
other hydrodynamic interactions \cite{reviewchaos} can all produce
chaotic behavior.  In this manuscript, we look at chaotic behaviors
involving particles in a horizontal drum which cascade
under the influence of a constant external force (i.e. gravity),
as first studied by Mullin {\it et al.} \cite{drumchaos}.

In the study of classical Newtonian chaotic systems, there are
several heavily studied systems which serve as examples to which
other systems can be compared.  Simple experiments such as a forced
pendulum \cite{chaospendulum} are easily constructed laboratory
experiments which can be studied in depth to better understand
low dimensional chaotic behavior.
Fluid systems, while containing many
more degrees of freedom, also exhibit several geometrically
simple experiments which can produce interesting results,
giving us insight into the character of chaotic fluid dynamical
systems \cite{vanhook97}.  Relevant to this manuscript, consider
non-Brownian particles sedimenting under the influence of gravity.
When many particles sediment, interesting swirling motions are seen
\cite{segre97}.  When only a few particles sediment interesting
effects are still seen, including chaotic behavior
\cite{expersed,stokeslet}.  Specifically, if three particles are
released close to each other, two of the particles eventually
pair up and travel together, while the third particle moves on its
own isolated trajectory.  Intriguingly, this pairing is sensitive
to initial conditions.  A slight change in the initial particle
positions can lead to a switch of which particle is isolated.

\begin{figure}
\centerline{
\epsfxsize=0.6\columnwidth
\epsffile{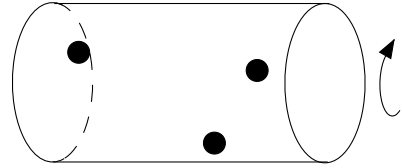}}
\smallskip
\caption[Rotating Drum]{Sketch of rotating drum filled with viscous
fluid and three beads.  These beads are dragged upwards by the
front wall of the drum until gravity pulls them away from the wall
and they fall through the fluid.  (Based on
Ref.~\cite{drumchaos}.)
}
\label{fig.drumcartoon}
\end{figure}

These simple sedimentation experiments have only transient
chaotic behavior which is difficult to study experimentally.
By considering motion in a rotating drum, particles can
interact repeatedly, leading to longer-lasting and potentially
persistent chaotic behaviors.  In a rotating drum partially
filled with a liquid and particles, segregation eventually
occurs \cite{acrivos99,acrivos99b}.  In a purely granular
system (no added liquid), similar segregation is also seen
\cite{choo97,choo98,williams76,bridgwater76,hill94,hill97}.
A simpler system was devised by Mullin {\it et al.}, in
analogy with the three-particle sedimentation simulations of
Ref.~\cite{stokeslet}.  They studied a hollow cylinder, filled
with glycerine, oriented horizontally, and rotated at various
angular speeds $\omega$ \cite{drumchaos}.  Three large non-Brownian
spheres are placed in this drum.  This system serves as a version
of a simple sedimentation experiment, with the constraint that
the rotation of the drum forces the three beads to remain close
together, continually interacting.  Figure \ref{fig.drumcartoon}
shows a sketch of this experiment.  When the drum rotates,
these beads can undergo a periodic cascade in the vertical direction.
As they cascade, the beads experience long-range interactions due to
the fluid.  Mullin {\it et al.} found several different behaviors
depending on $\omega$.  Simple behaviors included states where the
three particles remain well-separated and cascade independently of
each other.  More complex behaviors were observed, such as a state
described as ``chaotic'' where the particles move back and forth
horizontally along the tube as they continue to tumble vertically.
These horizontal motions were slow, erratic, and unpredictable.
In some cases two particles even momentarily collide before
withdrawing.  No chaotic behavior was observed when there were
only one or two beads.

Motivated by Ref.~\cite{drumchaos}, we study a similar experimental
system of a rotating drum with three spheres.  The benefit of this
experiment is that it can be stably run for very long periods of
time, thus allowing us to study the motions of the spheres over a
long period of time and describe the chaotic states in great
detail (which was not done in Ref.~\cite{drumchaos}).  We find the
states seen in Ref.~\cite{drumchaos} as well as three new states.
Additionally, we characterize the chaotic states, finding that
different chaotic states at different rotation rates $\omega$
often have widely different behaviors, despite a superficial
similarity.  We thus demonstrate and characterize a simple system
which possesses a wide range of nontrivial behaviors.

\section{Experimental Methods}
\label{apparatus}

\subsection{The Drum}
\label{ss.drum}

The experimental apparatus is a $l_c =25$~cm long horizontally
oriented sealed drum, with an inner radius $r_c = 5.72$~cm. These
dimensions are similar to those used in Ref.~\cite{drumchaos}
($l_c = 25$~cm and $r_c=5.9$~cm).  The main body
of the drum is constructed from a section of acrylic glass pipe
with a wall thickness of 1~cm.  To each end of this drum,
waterproof threaded aluminum caps are fitted.  Each cap has a
shaft attached via an adjustable mounting, so that the shaft can
be carefully centered within the cap.  These shafts attach to
a stand via two bearings, which allow the drum to rotate freely
about the horizontal axis.  The base of the drum stand contains
four adjustment screws for leveling the apparatus.

On one of the drum shafts, a pulley is mounted, and connected via a
belt to a Dayton 1/2 HP 3-phase A/C motor driven by a Fuji AF-300
controller.  This allows the drum to rotate on its axis at a
variable rotation rate $\omega$ from $5-13$~rad/s.
Due to the belt drive connection, the actual motor rotation rate
is potentially different from that displayed on the control box,
so the rotation rate of the drum is measured independently using
a Pasco PS-2120 rotary motion sensor connected to a computer.
Measurements taken over four hours show the
rotation rate to be stable to within $1\%$.  As all experiments
are started from rest, we also measure the time for the drum to
spin up to within $1\%$ of its final velocity, and find typical
spin-up times to be on the order of tens of seconds.

The drum is filled with $99.5$\% pure glycerol from Sigma-Aldrich,
and three 440c stainless steel ball bearings purchased from
Winstead Precision Ball Company, each of which has a diameter of
$2 r_b=1.59$~cm, density $\rho_b = 7.65 \times 10^3$~kg/m$^3$,
and mass of $16.1$ grams.  When immersed in glycerol ($\rho_f =
1.26 \times 10^3$ kg/m$^2$), the beads have an apparent buoyant
weight of $0.131$~N each.


To maintain constant viscosity, we control
the temperature of the fluid within the drum \cite{shankar94}.
To do so, we immerse the drum in a tank of water, in which a copper
heat exchanger has been placed.  We connect this heat exchanger
to a Thermo NESLAB RTE-7 digital refrigerated/heated bath, which
is maintained at 25$^\circ$~C in all experiments.  The rotation
of the drum provides sufficient mixing to allow the water within
the tank to be maintained at $24.7{\pm}0.1^\circ$~C, as measured
by a digital thermometer, independently for each experiment.
This results in a measured kinematic viscosity $\nu = 7.69$
Stokes (as compared with 9.36 Stokes for the fluid used in
Ref.~\cite{drumchaos}).

\subsection{Data Collection}
\label{ss.imaging}

To image and track the particles, we use a Pixelink PL-B741F 1.3
megapixel firewire monochromatic camera connected to a personal
computer running Windows XP.  A Monarch Instruments Nova-Strobe DAX
is used to light the particles from behind, which minimizes the
motion blur.  An opaque screen surrounds the entire apparatus to
block ambient light.  Compressed movies, lasting up to six hours
in duration, were captured using a custom application written in
C++ and analyzed with a particle tracking algorithm implemented
in Matlab.  We can locate particle positions with a resolution of
$\pm 0.25 r_b$, limited by slight optical distortions.

For each experiment, we initialize the particle positions
by setting the drum rotation rate $\omega$ to that of a known
chaotic state, and stopping the drum when the three beads are
distributed equidistant from one another, with approximately 3.5
particle diameters spacing between the beads.  The exact rotation
rate is not important, as it is simply used as a tool to position
the particles.  Once the drum is stopped and enough time allowed to
elapse for any fluid motion to cease ($\sim 10$ min), we start
the rotation of the motor at the desired rotation rate $\omega$, and
immediately begin recording video.  Videos are streamed directly
to the PC hard drive into a compressed Xvid MPEG-4 AVI file.
Video files are then post-processed using Matlab.  Note that all
experiments discussed follow this protocol, starting the drum
from rest, setting the speed on the motor for the desired
$\omega$, and then turning the motor on.  In
particular, we do not examine hysteretic effects (although we
speculate that there likely are some hysteretic effects, as
discussed below).

\subsection{Sphere Motion and Nondimensional Numbers}
\label{s.fluidbeads}

When the drum first begins to rotate, there is a transient state
where the fluid is not yet equilibrated to the new rotation rate.
This can be estimated by the Ekman pumping time,
\begin{equation}
\tau_E = \frac{l_c}{ \sqrt{\nu \omega}},
\end{equation}
based on the cylinder length $l_c$, viscosity $\nu$, and final
rotation rate $\omega$ \cite{spinup}.  For our experiment, with
$\omega \sim 5-13$ rad/s, we have 
$\tau_E \sim 2.4-3.7$ s.  This is faster
than the time needed for the motor to reach full speed (tens of
seconds as noted above), and so the motor is the limiting factor
in reaching the steady state.  This implies that the behavior seen
 in our measurements, lasting several hours in duration, is not
dependent on effects from the initial spin-up of the drum fluid.
Observations of tracer material within the fluid confirm the short
spin-up time.
In Sec~\ref{s.fourier}, we will show that typical time scales for
horizontal motion of the spheres ($x$ direction) are $O$(100~s), two
orders of magnitude removed from the Ekman time $\tau_E \sim 3$~s.
The particle turnover time (the time for a single cascade to occur)
is of order $\sim 1$~s.


\begin{figure}[htp]
\centering
\includegraphics[width=\columnwidth]{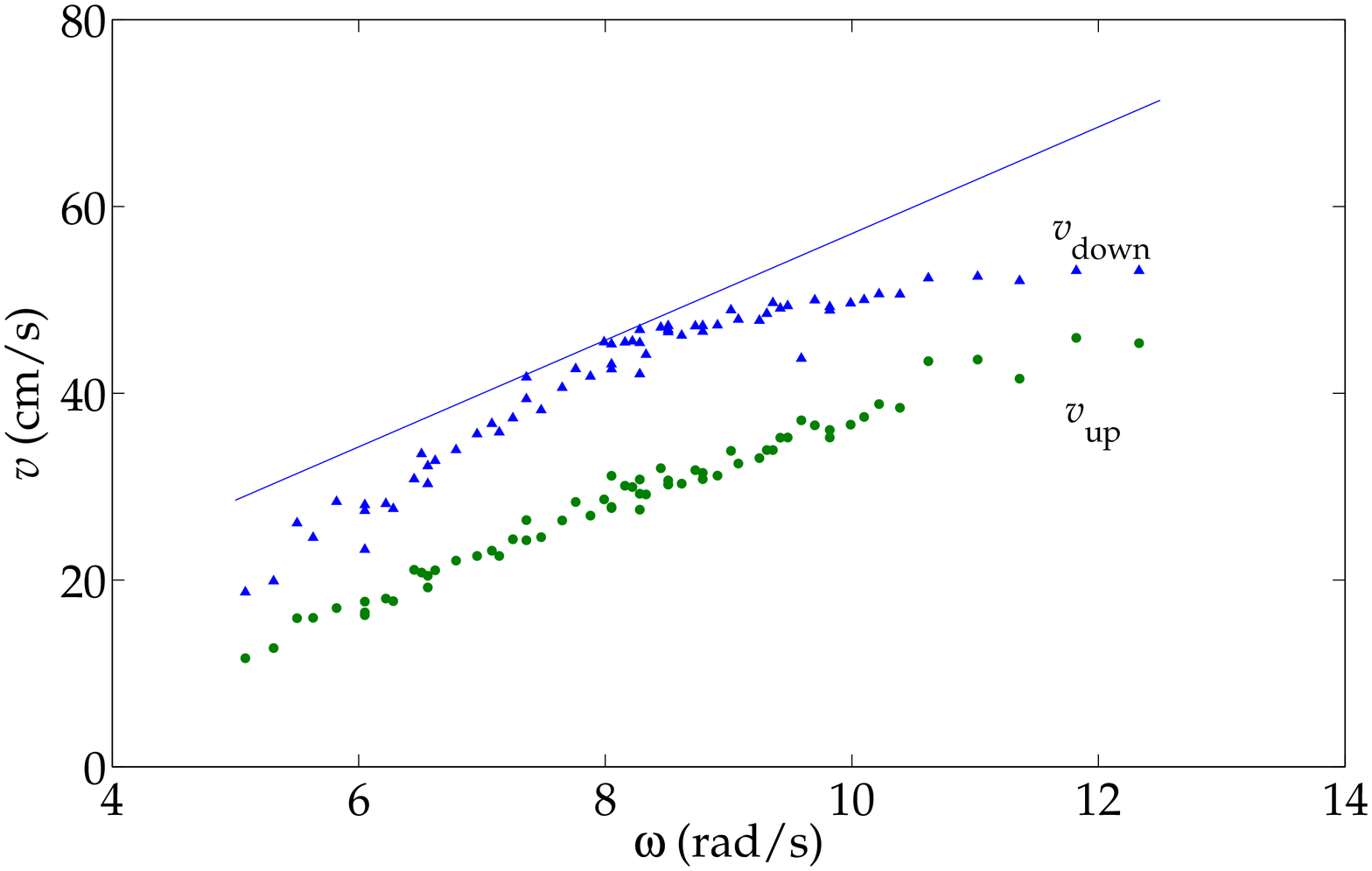}
\caption{
(Color online)
Cascade speeds $v_{up}$ and $v_{down}$ plotted versus $\omega$.
The solid line represents $v=\omega r_c$, showing that $\omega r_c$
is an approximate upper bound for the cascade speed.}
\label{fig.cascadev}
\end{figure} 

When we are
within the cascade regime, the beads repeatedly cascade in the
$y-z$ plane, being dragged up the front wall of the cylinder,
then falling away from it.  We measure the average speed of the
spheres as they rise and fall, and plot these speeds as a function
of $\omega$ in Fig.~\ref{fig.cascadev}.  The solid line drawn at
$v=\omega r_c$ is an approximate upper limit to the characteristic
speed $U$ of the particles.  Using this, we can compute the particle
Reynolds number
$Re = UL / \nu$, which determines the contribution of fluid
inertial effects relative to viscous drag.
Using the bead radius to set the characteristic length scale
$L=r_b$, the Reynolds number is \cite{drumchaos}:
\begin{equation}
Re = \frac{\omega r_c r_b}{\nu}    .
\end{equation}
Noting that, for this experiment, $r_c$, $r_b$, and $\nu$ are all fixed, we can write a simple linear conversion
\begin{equation}
Re= 0.591 \omega \sim 2-8 
\end{equation}
for the range of $\omega$ we study.  These low values of Re
correspond to a laminar flow where both viscosity and fluid inertia
play a role in the fluid flow.  Next, to quantify the importance
of viscous forces for influencing the horizontal motions of the
beads, we consider the typical viscous damping time $\tau_{\nu}
\sim L^2/\nu$.  Here we use $L = l_c$, to quantify interactions
across the length of the drum, and find $\tau_{\nu} \sim 80$~s,
on the same order as the particle interaction time $\tau_{col}
\sim 100$ s.

If we consider the ratio of inertial forces for the particles
compared to the fluid, we can typify the relative contributions
by comparing the density ratio $\rho_b / \rho_f \sim 6.2$, which
implies that fluid inertia will have less influence upon the bead
trajectories than the Reynolds number might otherwise imply.  Due to
their large inertia relative to the fluid, the beads will not tend
to follow fluid streamlines exactly, but rather interact via drag.

Finally, we consider the Galilei number to quantify the ratio
of gravitational to viscous forces.  The force due to gravity
is $F_g=(4/3) \pi r_b^3 \Delta \rho g$, and the viscous drag is
defined as the Stokes drag $F_{\nu}=6 \pi \nu r_b U$.  Using these
we define the Galilei number as \cite{masstransfer}
\begin{equation}
Ga = \frac{F_g}{F_{\nu}} = \frac{2 r_b^2 \Delta \rho g}{9 \nu U } \sim 1.3 - 3.9
\end{equation}
with $Ga = 3.9$ for $\omega = 5$ rad/s and $Ga = 1.3$ for $\omega
= 15$ rad/s, using the typical velocity scale as $U=\omega r_c$.
This shows that the influence due to gravity is always comparable
to that due to viscous drag, which is not surprising.  At lower
rotation rates, the force due to gravity is proportionally larger,
meaning the particles will not be lifted up as high in the cylinder
($y$ direction) before falling back down; this is indeed what
we observe.  Likewise at higher rotation rates, the forces due to
gravity and viscous drag are equal in magnitude, and the cascading
motion carries the spheres further upward in $y$.

\section{Results}
\label{s.results}

\subsection{Trajectories}
\label{ss.phasedescription}

\begin{figure}[thp]
\centering
\includegraphics[width=0.8\columnwidth]{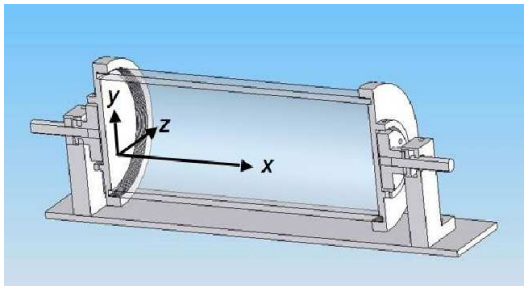}
\caption{(Color online)  The axes are defined such that $x$
represents the horizontal direction, $y$ is the vertical, and $z$
is the depth away from the front edge of the drum.  Note that, with
our current apparatus, there is no way to measure $z$ directly.}
\label{fig.axes}
\end{figure}

In Ref.~\cite{drumchaos}, Mullin {\it et al.} describe three types
of behaviors for the three bead case.  At low Reynolds number (Re
$<$ 1.21), they observed fixed-point behavior, where the beads
were completely independent of each other.  At 1.21 $<$ Re $<$
2.12, the beads underwent cascading motion in the $y-z$ plane
with the $x$ positions fixed, with the axes defined as drawn
in Fig.~\ref{fig.axes}.  It was noted that as they cascaded,
the outer beads were stably out of phase with each other, and
the middle bead was at an intermediate position between the two.
At Re $=$ 2.12 there was a reversible transition to a chaotic regime
where the particles started to wander erratically in the horizontal
($x$) direction while still cascading in the $y-z$ plane.  At Re
$=$ 4.53, there was a transition to what Ref.~\cite{drumchaos}
described as solid body motion.  In all
cases, the motion in the $y$ direction is always the simple
cascading motion, and the motion in $x$ is nontrivial; thus, like
Ref.~\cite{drumchaos}, we will focus on the $x$ motion for our
analysis.

\begin{figure}[htp]
\centering
\includegraphics[width=3.0in]{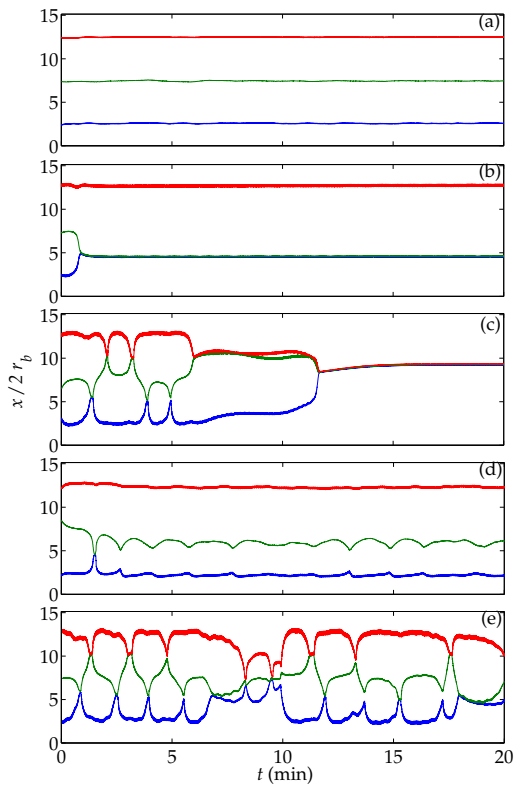}
\caption{
(Color online)  Typical examples of trajectories corresponding to
the five regimes we observe.
(a) Periodic trajectory with $\omega=5.31$rad/s.
(b) Stable doublet state with $\omega=7.14$ rad/s.
(c) Stable triplet state with $\omega=7.88$ rad/s
(d) Biased chaotic state with $\omega=6.56$ rad/s.  In such
states, the center bead strongly interacts with only one of the
outer beads.
(e) Fully chaotic state with $\omega=8.51$ rad/s.  For all
figures, the varying thickness of the lines is a result of particle
position uncertainty due to optical distortion from the curved
drum walls.  This distortion is exaggerated due to parallax near
the end caps, and minimized in the center of the drum.}
\label{fig.trajectories}
\end{figure}

In our experiment, we observed five distinct types of behavior
within the cascade regime described by Ref.~\cite{drumchaos},
three of which are new behaviors.  At the lowest
rotation rates, the three beads simply cascade in the $y-z$ plane
with no significant motion in the horizontal ($x$) direction,
corresponding to the cascading observations of
Ref.~\cite{drumchaos}.
Figure~\ref{fig.trajectories}(a) shows a typical example of
this periodic trajectory, at a rotation $\omega=5.31$ rad/s.
Each bead appears to move isolated from the influence of the
other beads.
As rotation rate is increased beyond this initial simple periodic
regime, there are several observed types of trajectories, depending
on rotation rate.  The two simplest of these are also periodic,
but to differentiate their unique behaviors we have labeled
them as doublet [Fig.~\ref{fig.trajectories}(b)] and triplet
[Fig.~\ref{fig.trajectories}(c)] states.  In the original study
\cite{drumchaos}, neither the doublet nor triplet states were
mentioned.

In the doublet state, two of the beads will lock together so that
they are cascading in one another's wakes.   The determination
of which two beads will tend to pair up is a result of initial
conditions, and not a systematic trend in the experimental
apparatus.  Simply stopping and restarting the drum at the same
rotation rate can sometimes switch which two beads will form a pair.

In the triplet state, all three beads come together and cascade in
line with one another, in a similar fashion to the doublet state.
The three beads can be stacked on top of each other, touching,
or they can be spaced out within the drum, following each others'
wakes without touching, depending on whether the beads are closer
to in phase or out of phase as they approach one another.  Both the
doublet and triplet states are stable configurations and have been
tested to remain locked for periods exceeding twenty-four hours
in duration.

At certain rotation rates, the beads will wander chaotically in
the $x$ direction.  For chaotic trajectories with a low enough
$\omega$, there will be a bias to one side of the drum or the other.
This biased chaotic trajectory is illustrated by a typical example
as shown in Fig.~\ref{fig.trajectories}(d).  Two of the three beads
will tend to repeatedly approach and interact with one another,
while the third bead will remain segregated at the far end of
the drum.  This third bead still experiences long range forces
from the other two beads, and can be seen to move in phase with
the collisions of the other two.  The determination of which two
beads will tend to pair up is again seemingly a result of initial
conditions, analogous to the doublet state.  Like the doublet
and triplet states, this biased chaotic behavior was not seen
previously \cite{drumchaos}.

For higher $\omega$ chaotic states, the beads explore a more rich
set of interactions, where they wander somewhat erratically in
the horizontal direction, occasionally even colliding with each
other.  The collisions observed include
pair collisions (left-middle and
right-middle) as well as triplet-type collisions where all three
beads come together.  Figure~\ref{fig.trajectories}(e) shows a typical
example of this behavior, which we call the fully chaotic state,
similar to the behaviors illustrated in Ref.~\cite{drumchaos}.
In pair collisions, beads can be either in phase or out of phase
with one another (in the cascading direction).
In phase collisions are more direct, with the
beads immediately colliding and moving away, while out of phase
collisions often involve the beads cascading over one another
several times before colliding.  Triplet-type collisions generally
involve two beads cascading over one another while a third bead
approaches and collides with them.  Note that when we have out of
phase collisions, the two beads have identical $x$ positions for
a while; our data acquisition rate is not fast enough to
carefully follow their motion in $y$, and we cannot
distinguish between the beads at that point.  Thus, it is likely
that in some cases, the two beads exchange places, but we cannot
detect this.  For example, in Fig.~\ref{fig.trajectories} the middle bead is
always drawn with the same color, but it is important to
recognize that it is quite possible that the identity of this
bead changes at collisions.

\subsection{Phase Diagram}
\label{ss.phasediagram}

\begin{figure}[bhtp]
 \centering
 \includegraphics[width=1.0\columnwidth]{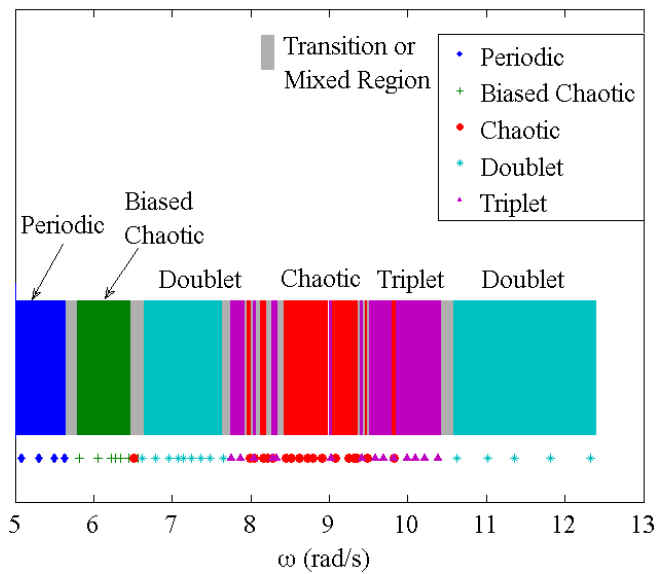}
 \caption{
(Color online)
Phase diagram showing the behaviors observed at various rotation
rates.  The symbols below the bar indicate individual observations.
 }
 \label{fig.phasediag}
 \end{figure}

To probe the dependence of the particles' behavior on rotation rate,
we recorded sixty-eight videos at rotation rates ranging from 5.1 to
12.3 radians per second.  An analysis of these videos allows us to
map out a phase diagram as shown in Fig.~\ref{fig.phasediag}.  The
colored blocks denote different regimes, and gray blocks represent
regimes where there is some overlap of behavior, or transitions
between two regimes.  The width of these transition blocks is due
to uncertainty both due to the measurements themselves, and also
to the discrete, digital motor control circuit, which limits the
resolution of $\omega$ to 1\% as noted in Sec.~\ref{apparatus}.

With increasing rotation rate $\omega$, the system undergoes
a phase transition from periodic to biased chaotic behavior
at $\omega \sim 5.7$, which corresponds to a Reynolds number
$\text{Re}\sim 3.4$, comparable to the transition from periodic
to cascading motion seen in Mullin {\it et al.}'s experiment at
$\text{Re}=2.12$ \cite{drumchaos}.  The difference in Re is perhaps
due to our different Galilei numbers; because their viscosity was
1.2 times larger than ours, their values for $Ga$ are smaller by
that same ratio.

In our observations, the biased chaotic regime is followed by a
long doublet regime.  After this doublet regime, we find a small
window of triplet behavior around $\omega \sim 7.8$ rad/s, which
begins a mixed region of behavior, consisting of slices of both
chaotic and triplet behavior, extending until $\omega \sim 10.4$.
For rotation rates higher than those at which we find triplet
behavior, we find reliable doublet trajectories.  For
high enough rotation rates, we should transition into the motion
Mullin {\it et al.} described as solid-body \cite{drumchaos}.
We do not probe this regime due to limitations of our motor driving
the rotating drum.

This phase diagram illustrates a rich landscape of interesting
regimes of particle behavior with a new level of detail.
Specifically, Ref.~\cite{drumchaos} identified only one simple, contiguous
block of chaotic behavior, while we have identified
multiple windows of periodic behavior embedded within large chaotic
regimes, as well as previously unidentified periodic behaviors.
Furthermore, the distinction between two different types of chaotic
behavior illustrates the complexity of the system.

\begin{figure}[bhtp]
\centering
\includegraphics[width=1.0\columnwidth]{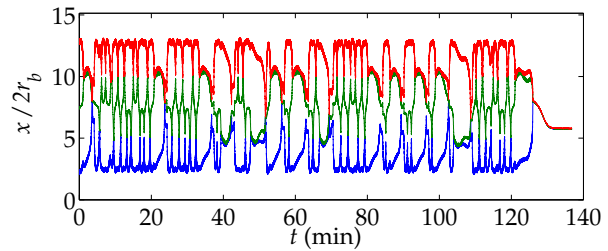}
\caption{
(Color online)
In this experiment, at $\omega = 8.05$ rad/s, the
particle follows a seemingly chaotic trajectory for $\sim 120$
min, but then settles into a stable triplet state. }

\label{fig.triptransient}
\end{figure}

There are two inter-related caveats to be considered when
discussing the phase diagram in Fig.~\ref{fig.phasediag}, transient
behavior and motor drift.  Transients can pose potential issues in
situations where the transients last longer than the duration of an
experiment.  In a given experiment, after the drum begins rotating,
the system takes some time to settle into its long-term behavior.
For example, in Fig.~\ref{fig.trajectories}(c), the particles move
back and forth across the drum, colliding several times before
finally coming together to form the triplet state at $T \sim
12$ min.  In this case, the time is small compared to typical
experimental durations ($T \sim 300$ min).  However, in other
experiments, such as that shown in Fig.~\ref{fig.triptransient},
transient behavior can persist for longer periods of time.  Here,
the trajectory is seemingly chaotic for $\sim 120$ min before
settling into a stable triplet configuration.

\begin{figure}[htp]
\centering
\includegraphics[width=1.0\columnwidth]{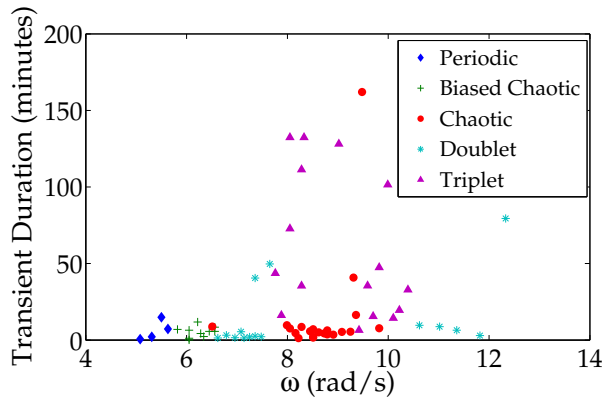}
\caption{
(Color online)
The approximate time taken for the initial transient behavior
to die out is obtained by eye and plotted versus rotation rate.
Symbols represent the long-term phase of each trajectory, after
its transients have died out.}
\label{fig.transients}
\end{figure}

To further explore the impact of these long
transient trajectories, we examine each trajectory
and manually determine an approximate transient duration.
Figure~\ref{fig.transients} shows this transient duration plotted
versus the rotation rate of the drum.  The symbols in the graph
represent the type of trajectory found after the transient behavior
has died out.

Many trajectories have relatively
short transient times, with transients rarely exceeding sixty
minutes in duration.
However, there are also trajectories which contain
much longer transient durations, with most of these long-duration
transient trajectories clustered around the transitions between
different phases.  One possible explanation for the long-lived
transients is the drum rotation rate, which as Sec.~\ref{ss.drum}
noted is stable to within $\sim 1\%$.
If we consider the transition around
$\omega = 8$ rad/s, we see that, for a given trajectory, $\omega$
could vary from $7.92$ to $8.08$ rad/s.
Thus a possible source for the long
transient behaviors is drift in rotation rate of the drum motor.
If we imagine small windows of triplet behavior within a chaotic
regime, a drum rotation rate which starts within the chaotic regime
could drift into the triplet regime, leading to a trajectory which
eventually ``finds'' the triplet state.  As already discussed,
the triplet state is very robust and stable, and thus once a
trajectory finds this state, it would be very difficult to break
out of it, even if the rotation rate wanders subsequently.  The fact that
long transients tend to cluster around the transitions between
regimes supports this hypothesis.

This issue of motor drift also has the potential to obscure
some detail in the phase diagram.  The motor control has finite
resolution in available rotation rates, and so there may be small
windows of behavior which we are unable to locate.  Similarly,
even if we did sample these windows, motor drift could take the
rotation rate out of a window if it existed within a very narrow
range of rotation rates.

\subsection{Qualitative Fluid Behavior}
To qualitatively describe the fluid flow within the drum,
we added a small quantity of Kalliroscope rheological fluid to the
glycerol within the drum.  Kalliroscope is a water-based suspension
of microscopic crystalline platelets.  When placed within a moving
fluid, the platelets tend to align such that their long axis is
parallel to the plane of shear.  Thus, the platelets will reflect
different amounts of ambient light depending on the local flow of
the fluid.  This allows us to visualize the flow behavior in each
of the states.

\begin{figure}
\centering
(a)\includegraphics[width=2.8in]{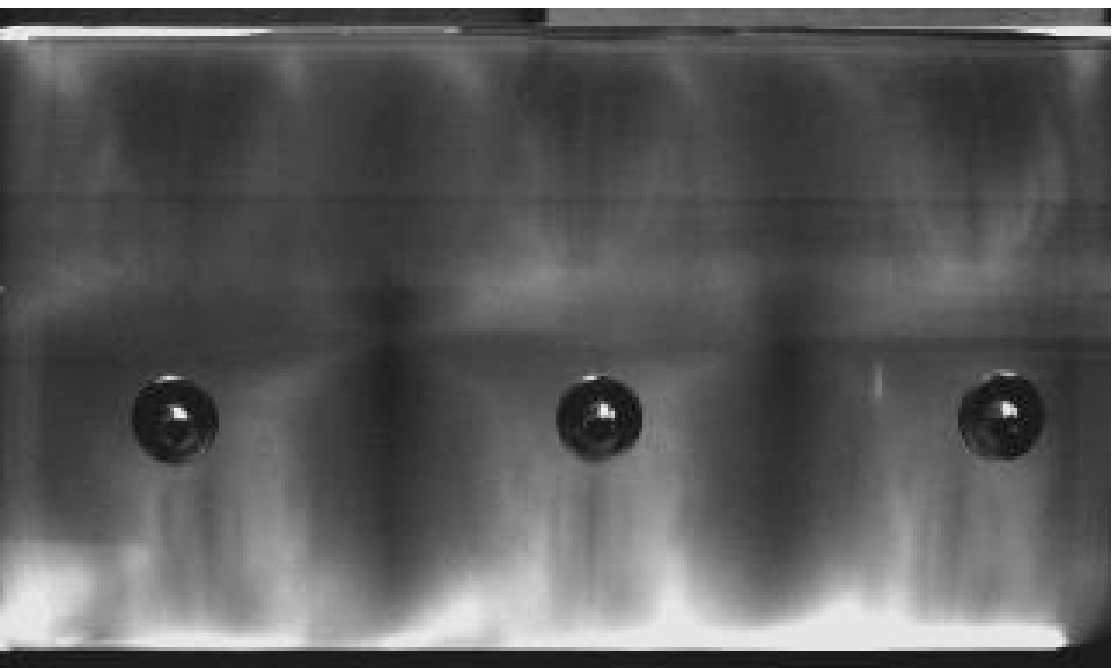}
(b)\includegraphics[width=2.8in]{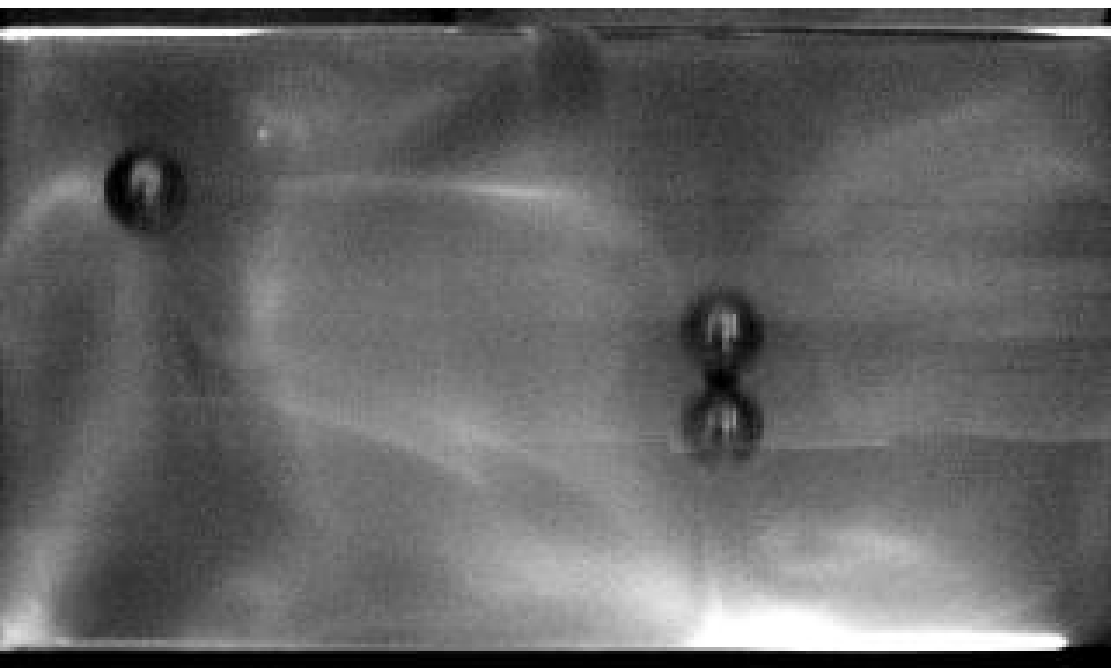}
(c)\includegraphics[width=2.8in]{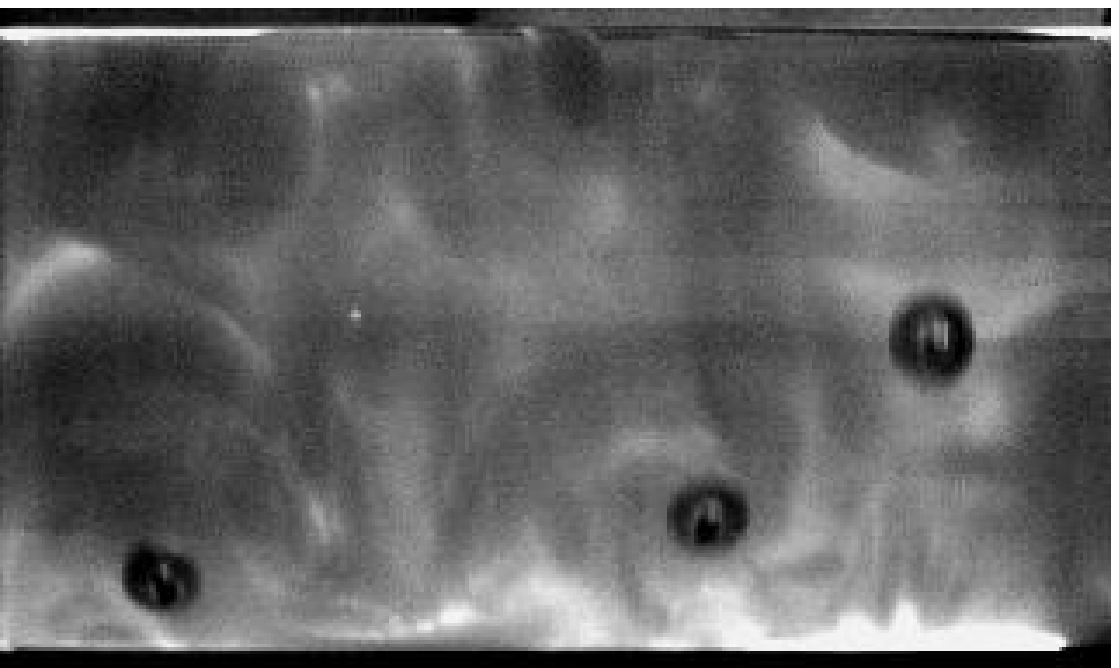}
\caption{Kalliroscope images show the fluid behavior in various
phases: (a) periodic ($\omega=5.31$~rad/s), (b) doublet
($\omega=7.14$~rad/s), (c) chaotic ($\omega=8.51$~rad/s).}
\label{f.kall}
\end{figure}

Within the periodic regime, the three beads each have a well
defined wake, which is bounded on each side by swirling,
vortex-like behavior rotating about an axis that extends in
the radial direction, as seen in Fig.~\ref{f.kall}(a).  At the
midpoints between each pair of particles, there are well defined
shear planes which span the entire height of the drum.

In the doublet regime, the two beads which are paired up form a wake
which keeps them aligned with one another.  This wake is bounded on
each side by vortex-like regions where there is swirling fluid flow,
as shown in Fig.~\ref{f.kall}(b).  The single bead, well-separated
near the far end of the drum, also has a well defined wake, but
there is significantly less vortex-like behavior in the fluid.

In the triplet regime , there is one strong wake in which all three
beads cascade (not shown).  There is a large amount of vortex-like
swirling that bounds this wake and likely leads to the observed
stability of the triplet state.

The flow within the biased chaotic regime (not shown) appears
similar to that within the periodic regime, except when the
beads collide.  As the beads approach a collision, their wakes
overlap and partially merge.  At the same time, as the beads
are approaching one another, the shear plane that separates them
oscillates with greater and greater amplitude, until it breaks up
as they approach.  After a collision, when the beads are moving
apart, the fluid to the outside undergoes a strong vortex-like
swirling until the beads are well separated.

Within the fully chaotic regime, the beads' wakes are often less
well defined and more difficult to identify, with large regions of
complicated fluid flow, as shown in Fig.~\ref{f.kall}(c).  However,
when the beads are well separated, their wakes are evident, with the
wakes becoming mixed and obscured as the beads approach one another.
The well defined shear planes seen separating the beads in previous
cases are not evident in the fully chaotic regime.

In all cases the visualization makes it clear that there is no
turbulence, in agreement with the low Reynolds number (Re $\sim
2-8$, see Sec.~\ref{s.fluidbeads}).

\subsection{Timescales}
\label{s.fourier}

From Fig.~\ref{fig.trajectories}(e), we note that the particles
spend the bulk of their time in a well-separated state
where the three particles are spaced far apart in the drum.
This configuration is similar to the stable configuration seen
in the periodic state shown in Fig.~\ref{fig.trajectories}(a).
Disturbances of the trajectories away from this well-separated
configuration are relatively short by comparison.  An interesting
question, then, is how much time the particles spend in this well
separated configuration.

A visual inspection of representative chaotic trajectories seems
to indicate a typical time between collisions of the particles.
For example, in Fig.~\ref{fig.trajectories}(d) many of the
collisions between particles occur roughly 1-2 minutes apart.
Fourier spectra of the $x$ trajectories are noisy and do not
depend in any obvious way on the drum rotation rate $\omega$.
These Fourier spectra give an indication of typical collision
timescales, with typical peak frequencies $f$ between $0.3-1.5$
cycles per minute, with large changes in $f$ at nearly the same
$\omega$, corresponding to collision times between $0.6$ and $3.3$
minutes.

\begin{figure}
\centering
\includegraphics[width=2.5in]{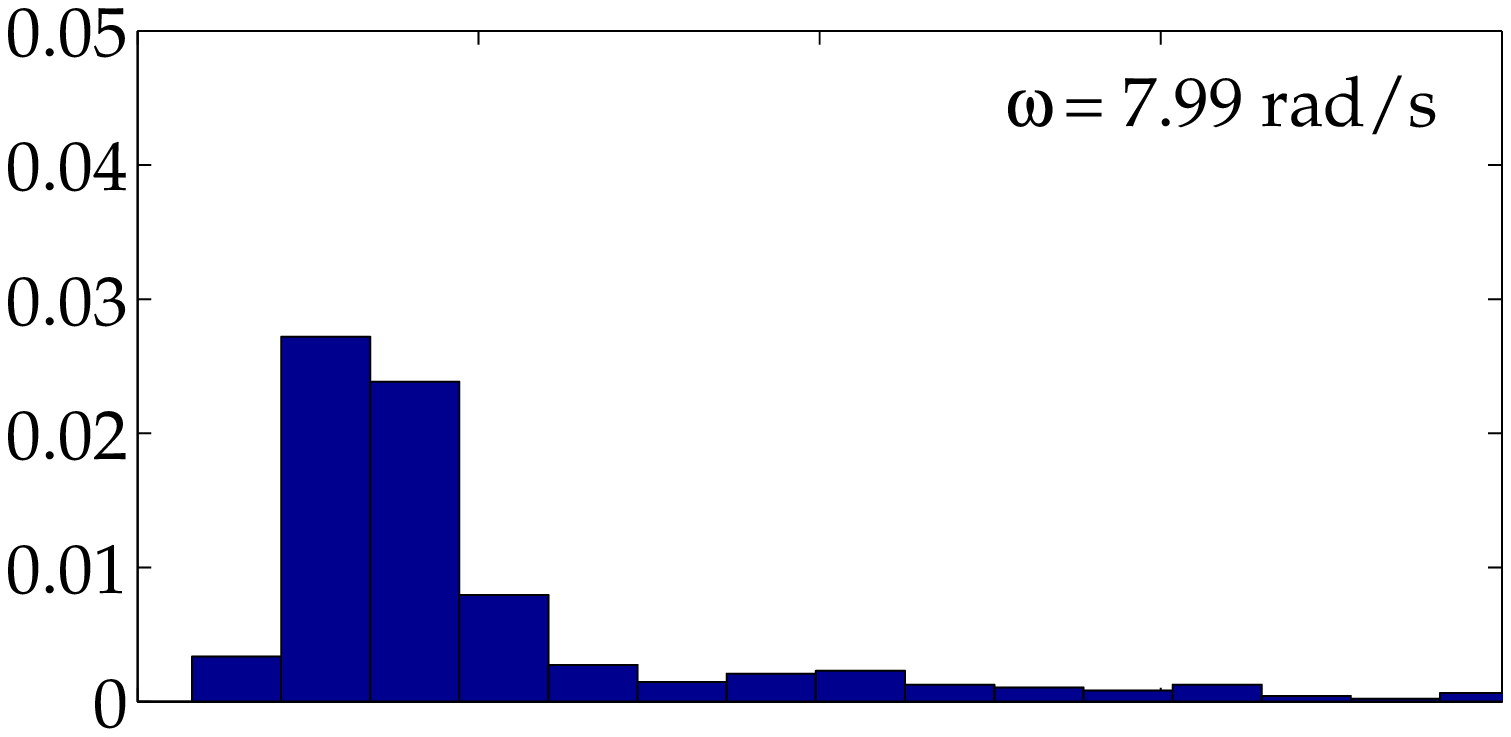}\\
\includegraphics[width=2.5in]{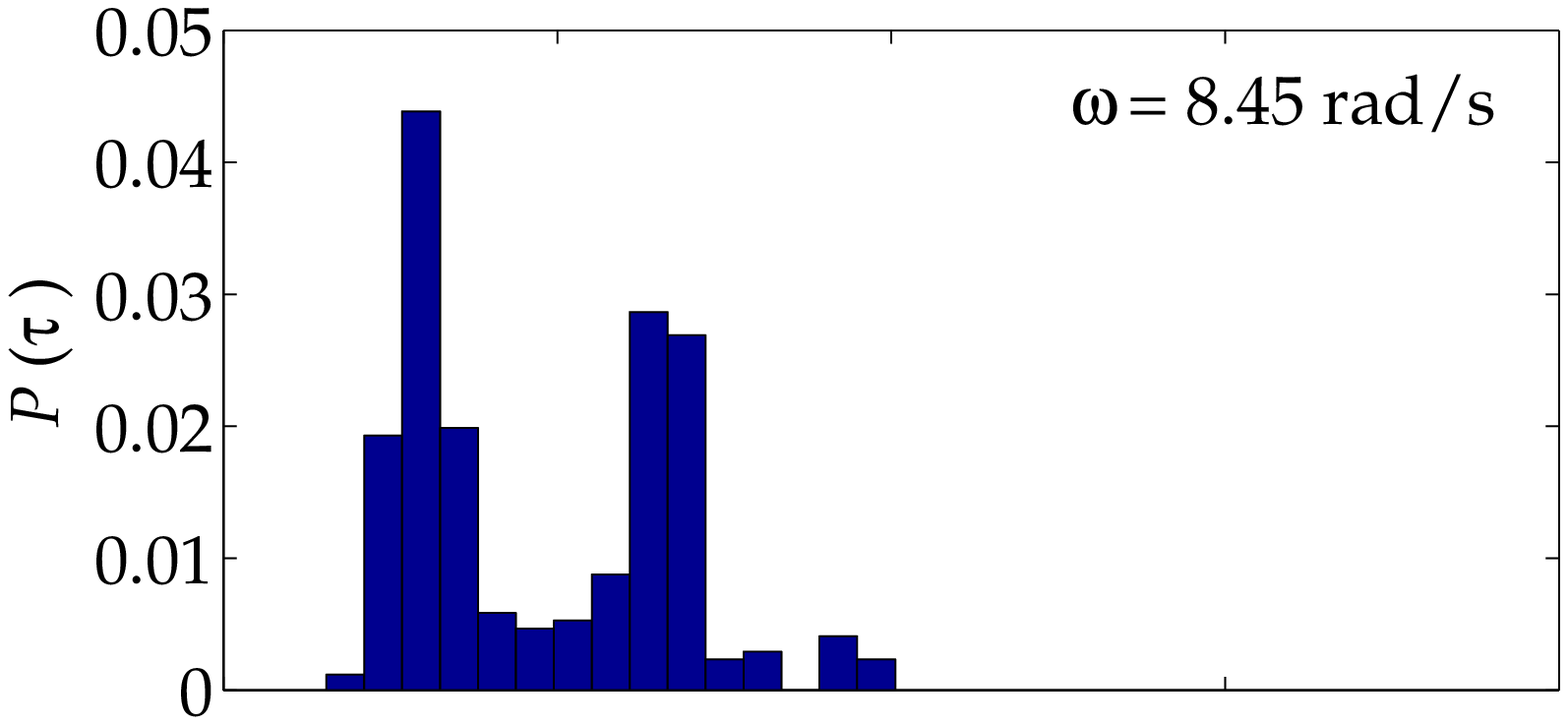}\\
\includegraphics[width=2.5in]{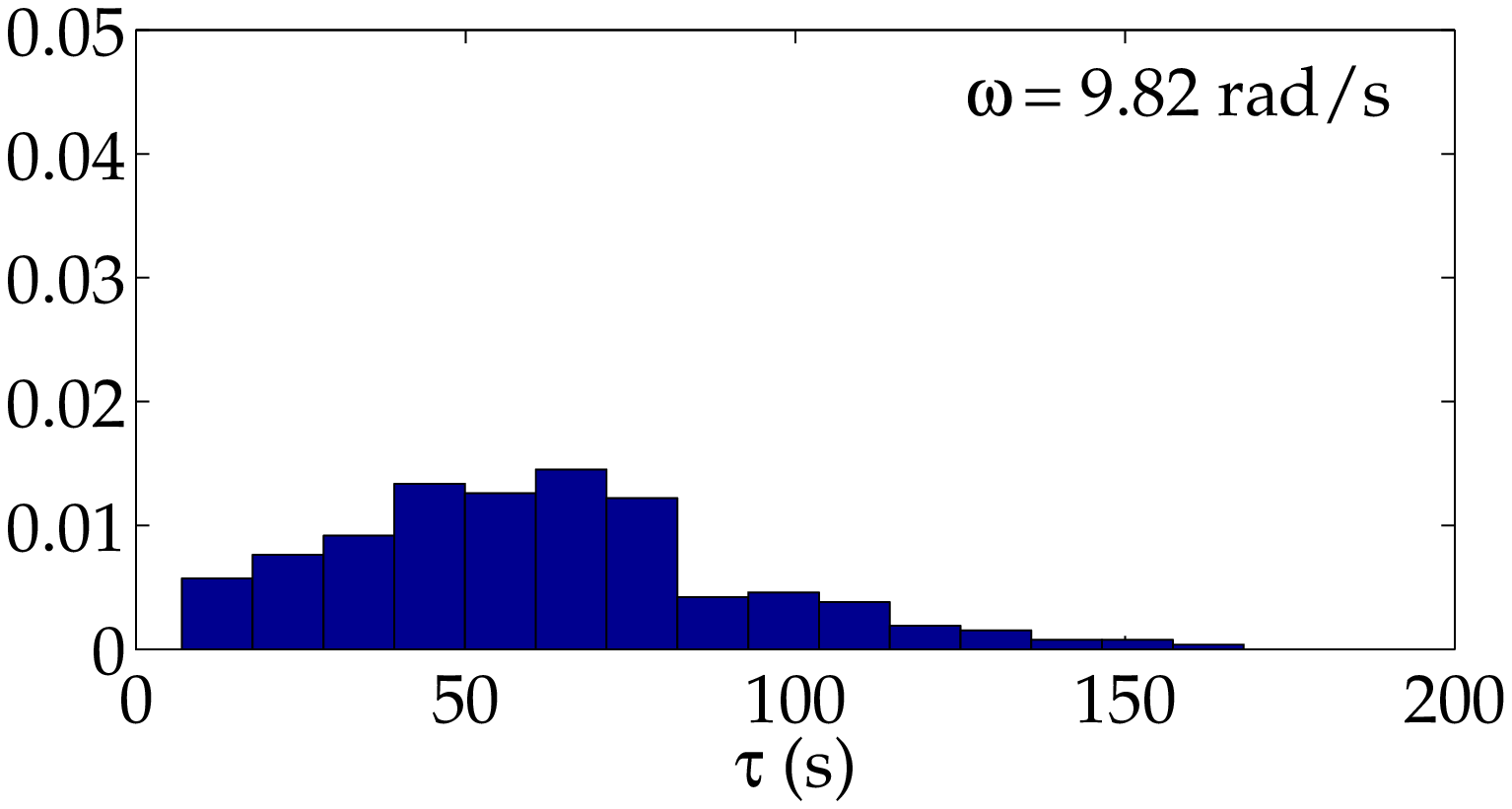}\\
\caption{
(Color online)
Distributions of time between ``collisions'' where two or three
beads come close together.  These all correspond to fully chaotic
states, with the drum rotation rates as indicated.
}
\label{fig.pkpkhists}
\end{figure}

To further examine the collision times, we analyzed the time between
particle interaction events.  To define these events, we note that
deviations from the well-separated configuration can be identified
by simply looking for local minima and maxima in the trajectory
$x_2$ of the middle bead.
Figure~\ref{fig.pkpkhists} shows the distribution of the time
$\tau$ between sequential peaks in $x_2$ for three representative
experiments within the chaotic regime.  There is no simple trend
in the shape of these graphs with varying $\omega$.  Overall,
the distributions of times are all broad with standard deviations
comparable to their means, reflecting that the particle trajectories
are unpredictable.  That is, particles spend a significant time in
well-separated positions, and then begin to come together for a
collision event after a variable amount of time.  Investigation
of sequential pairs of dwell times, $\tau_i$, $\tau_{i+1}$,
showed no structure, further implying unpredictability.  The state at
$\omega=8.45$~rad/s shows a bimodal distribution, which we observed
in only two out of the twenty chaotic states; the other state
was at $\omega=8.91$~rad/s, with several uni-modal distributions
observed at intermediate values of $\omega$.

\subsection{Reduced Dimensionality}

\begin{figure}[htp]
\centering
\includegraphics[width=\columnwidth]{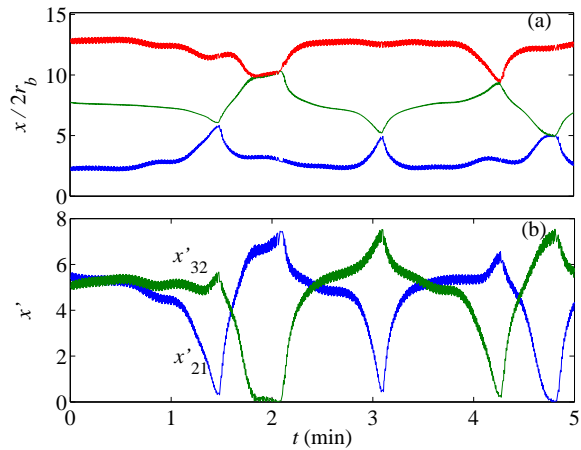}
\caption{
(Color online)
(a) Chaotic particle trajectories for an experiment
with $\omega = 8.79$ rad/s.  
Recall that we cannot distinguish the spheres from
one another, so that by definition $x_3 \geq x_2 \geq x_1$.
(b) The same trajectories in
reduced coordinates.  The minima in these
simplified trajectories indicate points where particle pairs
approach one another.
}
\label{fig.x21x32}
\end{figure}

To this point, all analysis has focused on the horizontal ($x$)
direction trajectories, and neglected cascading in the $y-z$ plane.
To further simplify the number of variables used in the
data analysis, we sought a reduced dimensionality set of variables
which still contains the interesting behavior of the system.
We note that, once trajectories have settled into their long-term
behaviors, and the transients have died out, the center of mass
of the system is constant within the combined noise in the three
particle trajectories.  This implies that the absolute positions of
the particles are not needed to capture the interesting behavior
of the system, and we can use a reduced dimensionality to study
the behavior.  Specifically we use the distances of each outer
bead from the center bead:

\begin{eqnarray}
x'_{21} = |x_2-x_1| / 2r_b \\
x'_{32} = |x_3-x_2| / 2r_b   .
\end{eqnarray}
Figure~\ref{fig.x21x32} shows a sample 
trajectory comparing the original coordinates with the reduced
coordinates.

As shown in Fig.~\ref{fig.x21x32}, there are certain configurations
in which the system spends more time, and other configurations
that are only visited briefly.  Configurations with the three beads
well separated appear to be most common, while situations where the
beads are close together are shorter-lived.  To quantify
this, we plot a two dimensional histogram of the configurations,
which gives us a way to visualize the relative amount of time each
particle spends in various regions of phase space.  Histograms are
plotted with a logarithmic intensity, so rare events can still
be identified.
In each experiment, we visually inspect the data set and remove
any obvious initial transient behavior manually before analysis.
For example, the analysis of a triplet data set only includes
the time after the three beads have lined up.

\begin{figure}[!htp]
\centering
\includegraphics[width=2.8in]{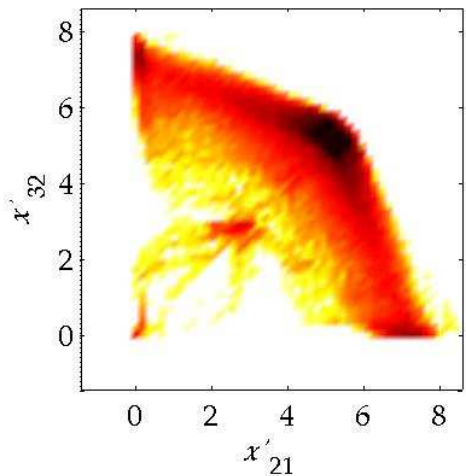}
\caption{
(Color online)
A typical two dimensional histogram for a chaotic trajectory
at $\omega = 8.79$ rad/s.  The axes represent
the distances between the pairs of particles, normalized by the
particle diameter, and color represents the number of points that
were counted in each bin.  Darker colors indicate more prevalent
configurations.
This experiment
has a calculated entropy $S=6.33$.}
\label{fig.examplehist}
\end{figure}

Figure~\ref{fig.examplehist} shows an example of one of these
histograms for the same experiment shown in Fig.~\ref{fig.x21x32}.
Notice that the darkest red region is in the area around $x'_{21}
\approx x'_{32} \approx 5-6$, corresponding to a configuration
where the three beads are spread far apart, and spaced roughly
equidistantly.  There are also small clusters at  $x'_{21} \sim 7.5$
and $x'_{32} \sim 0$, and its mirror  $x'_{21} \sim 0$ and $x'_{32}
\sim 7.5$, which correspond to configurations where two beads are
close together, and the third bead is far away.  Finally, there
is another faint cluster at  $x'_{21} \sim 0$ and $x'_{32} \sim 0$,
corresponding to a state where all three beads are grouped together.
The faintness of this cluster implies that very little time is
spent in this configuration.

\begin{figure}[hbp]
\centering
\includegraphics[width=3.0in]{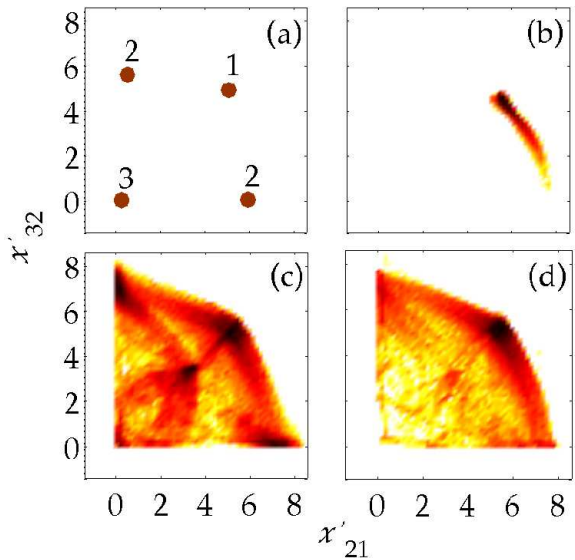}\label{fig.thehists}
\caption{
(Color online)
Histograms for each
phase of behavior clearly illustrate the amount of phase space
they explore.  (a) Schematic of typical areas occupied
by the the non-chaotic trajectories.   ``Periodic'' trajectories
such as shown in Fig.~\ref{fig.trajectories}(a) appear as a tight
cluster of points near position 1.  ``Doublet'' trajectories
reside at one of the two locations marked as position 2.
``Triplet'' trajectories have all the particles at the same $x$
position, and thus the histogram for these states is a tight
collection of points near the origin, marked as position 3.
(b) A histogram for a biased chaotic
trajectory at $\omega=6.05$~rad/s.
For this state, the entropy $S=3.52$.
(c,d) Histograms for chaotic trajectories at
$\omega= 9.25, 9.36$~rad/s, 
respectively.   The entropies of these states are $S=6.82, 6.49$.
}
\label{fig.hists}
\end{figure}

The histograms show how the different phases of the system explore
phase space.  Figure~\ref{fig.hists}(a) shows a schematic of typical
histograms for periodic, doublet, and triplet trajectories.
Periodic trajectories have well-separated beads and thus
the weight of the histogram stays concentrated at point 1.
Doublet states have a histogram with points clustered tightly
near one of the locations marked 2; for example, if beads 2 and
3 are together in the doublet state, then $x'_{23}=0$.  In the
triplet state, the three beads coincide in their $x$ coordinates
and thus the histogram is at the origin, where point 3 is shown.
Figure~\ref{fig.hists}(b) shows a biased chaotic trajectory, where
the beads spend time in the same region of phase space as the
periodic state, but also wander chaotically in the $x$ direction,
smearing the histogram in the direction of the bias (in this case,
beads 2 and 3 come close together).  In Fig.~\ref{fig.hists}(c) we
see a chaotic trajectory.  The beads explore the phase space around
the locations of periodic, doublet, and triplet states [points 1,
2, and 3 in Fig.~\ref{fig.hists}(a)].  Time is also spent in an
intermediate configuration ($x'_{12} \approx x'_{23} \approx 3.5$)
where the three beads are closer together than in the periodic
trajectory, but not clustered like in the triplet state. This is
in contrast to the chaotic state shown in Fig.~\ref{fig.hists}(d),
where the beads spend the bulk of their time well-separated, as
in the periodic state, with occasional doublet-like collisions.
The fact that these states occur at rotation rates which are very
similar ($\omega = 9.25, 9.36$ rad/s) illustrates how sensitive
the experiment is to the rotation rate.

\subsection{Entropy}

To quantitatively study the extent to which a given trajectory
explores phase space, we define a configurational entropy based
on these histograms.  If we first normalize a given histogram so
that the sum of all bin values is equal to unity, the histogram
will represent a probability distribution $P$ with matrix elements
$P_{ij}$.  We then define the entropy
\begin{equation}
S = -k \sum_{ij} P_{ij} \text{ln} P_{ij}
\hspace{0.1in},\hspace{0.1in} \text{with}\ k \equiv 1  .
\end{equation}

\begin{figure}[htp]
\centering
\includegraphics[width=3.5in]{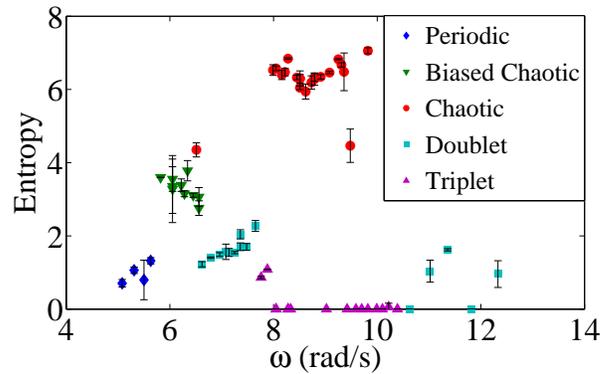}
\caption{(Color online)  The entropy for each rotation rate,
calculated from the 2-D histograms.}
\label{fig.entropy}
\end{figure}

This entropy value is calculated for each individual trajectory
and plotted versus rotation rate in Fig.~\ref{fig.entropy}.
The triplet states have the lowest entropy, followed by the
periodic, doublet, biased chaotic, and finally the fully chaotic
states with the highest entropy.
To quantify error in entropy measurements we split each trajectory
into two halves, and calculate the entropies $S_1$ and $S_2$
for each half independently. Then the error can by defined as:
\begin{equation}
 \sigma_S = |S_1 - S_2|       ,
\end{equation}
yielding the error bars shown in Fig.~\ref{fig.entropy}.

As could be expected, the entropy is lowest in the simplest
states:  the triplet, doublet, and periodic states.  Nonzero
values for the entropy correspond to slight wobbling of the
particles around their mean positions in each state.  At values
of $\omega$ corresponding to transitions between states,
Fig.~\ref{fig.entropy} shows sharp changes in the entropy.  For
example, the biased chaotic state has entropy values markedly
higher than the adjacent doublet states.  The fully chaotic
states have the highest entropy.  Within each regime, there can
be moderate fluctuations of the entropy values, with little
systematic dependence on $\omega$.

Much as the histograms provide a visual indication of the degree
to which a trajectory explores phase space, these entropy values
provide a qualitative measure of that exploration.  Low entropy
periodic, triplet, and doublet regimes do not explore phase space
much, as expected.  Chaotic behaviors, on the other hand, have large
entropies, corresponding to extensive exploration of phase space.
These entropies vary little by comparison to the large jumps between
regimes, indicating that the amount of phase space explored by
chaotic behaviors does not vary much with rotation rate.

The magnitude of these entropy values varies with the size of bins
used for the histograms.  We divide each axis into $60$
divisions, giving a total of $60^2$ bins, with each bin having a
width  $w \sim r_b/3$.  The qualitative results are unchanged
with $40^2$ or $80^2$ bins.

\section{Summary}

We have
studied a geometrically simple system containing three particles
moving within a fluid-filled rotating drum which yields a rich
and varied set of behaviors.  The phase diagram for this system
showed five types of behavior.  The first is a periodic regime
where the beads simply cascade in the $y-z$ plane.  The second is a
previously unreported biased chaotic regime where two of the beads
wander chaotically in the horizontal $x$ direction and collide with
one another.  The third is a doublet regime, where two beads pair
up and cascade on top of one another while leaving the third bead
behind.  There is also is a mixed chaotic regime spanning a wide
range of rotation rates, where the beads wander chaotically in the
horizontal direction, with all three beads interacting and mixing.
Within this mixed regime, there are small windows of rotation rates
which result in triplet behavior, where all three beads will line
up and cascade on top of one another.  Finally, we find a regime
where triplet behavior is the only type of trajectory seen.

The question of transient behavior deserves significant attention
and exploration.  We find chaotic states that are persistent
over many hours.  However, based on similarities between the
chaotic states and the transient behavior of the triplet states,
it is possible that the chaotic states could eventually fall into
a stable triplet state.  Our ``fastest'' observed chaotic states
have mean collision times of approximately 0.6~min, and our longest
observations are up to 360~min, so at most we have 600 collisions
observed in chaotic states without a transition to a triplet state,
thus at least showing that if these are transients that they are
very long-lived.

Another interesting question is that of dependence on initial
conditions.  Due to the way the particles were positioned within
the drum, it was difficult to control their exact starting
positions. However, the long-term statistics of the system's
behavior were reproducible, within the limits of motor drift and
transients.  

We have also tried preliminary experiments with a longer drum
(thus larger aspect ratio), and find that particles prefer doublet
or triplet states; we do not see long-lived chaotic states in
longer drums.  This suggests that a key to the chaotic states is
indeed the finite drum size that forces the beads to interact with
each other.  Overall, we have demonstrated a simple system with
complex and chaotic behavior, and furthermore shown that within
the previously observed chaotic regime (Ref.~\cite{drumchaos})
there is a richness of behavior with the ``degree'' of chaos
changing dramatically with only slight changes of rotation rate.

\begin{acknowledgments}

We thank M. Schatz and D. Borrero for guidance in the use of
Kalliroscope and many helpful discussions, and T. Mullin for
his inspiration of this project and helpful discussions.
This work was supported by the Emory University Graduate School
of Arts and Sciences and NSF DMR-0804174.

\end{acknowledgments}

\bibliography{latex/jdavidh}



\end{document}